\documentstyle[amssymb,aps,prd,multicols,epsf]{revtex}

\begin{document}
\title{A Further Analysis of a Cosmological Model with Quintessence and Scalar Dark Matter.}
\author{Tonatiuh Matos\thanks{E-mail: tmatos@fis.cinvestav.mx} and L. Arturo Ure\~{n}a-L\'{o}pez\thanks{E-mail: lurena@fis.cinvestav.mx}}
\address{Departamento de F\'{\i}sica, \\
Centro de Investigaci\'on y de Estudios Avanzados del IPN,\\
AP 14-740, 07000 M\'exico D.F., MEXICO.\\
}
\date{\today}
\maketitle

\begin{abstract}
We present the complete solution to a $95\%$ scalar field cosmological model in which the dark matter is modeled by a scalar field $\Phi $ with the scalar potential $V(\Phi )=V_{o}\left[ \cosh {(\lambda \,\sqrt{\kappa _{o}}\Phi )}-1\right] $ and the dark energy is modeled by a scalar field $\Psi $, endowed with the scalar potential $\tilde{V}(\Psi )=\tilde{V_{o}}\left[ \sinh {(\alpha \,\sqrt{\kappa _{o}}\Psi )}\right] ^{\beta }$. This model has only two free parameters, $\lambda $ and the equation of state $\omega _{\Psi }$. With these potentials, the fine tuning and the cosmic coincidence problems are ameliorated for both dark matter and dark energy and the models agrees with astronomical observations. For the scalar dark matter, we clarify the meaning of a scalar Jeans lenght and then the model predicts a suppression of the Mass Power Spectrum for small scales having a wave number $k > k_{min,\Phi}$, where $k_{min,\Phi} \simeq 4.5 \, h \, {\rm Mpc}^{-1}$ for $\lambda \simeq 20.28$. This last fact could help to explain the dearth of dwarf galaxies and the smoothness of galaxy core halos. From this, all parameters of the scalar dark matter potential are completely determined. The dark matter consists of an ultra-light particle, whose mass is $m_{\Phi} \simeq 1.1\times 10^{-23}\, {\rm eV} $ and all the success of the standard cold dark matter model is recovered. This implies that a scalar field could also be a good candidate as the dark matter of the Universe.

\end{abstract}

\draft
\pacs{PACS numbers: 98.80.-k, 95.35.+d}

\begin{multicols}{2} \narrowtext

\section{Introduction}

Observations of the luminosity-redshift relation of Ia Supernovae suggest that distant galaxies are moving slower than predicted by Hubble's law, implying an accelerated expansion of the Universe\cite{snia}. These observations open the possibility to the existence of an energy component in the Universe with a negative equation of state, $\omega <0$, being $p=\omega \rho $, called dark energy. This componente would be the currently dominant component in the Universe and its ratio relative to the whole energy would be $\Omega_\Lambda \sim 70\%$. The most simple model for this dark energy is a cosmological constant ($\Lambda $), in wich $\omega=-1$. The matter component $\Omega _{M} \sim 30 \%$ of the Universe decomposes itself in baryons, neutrinos, etc. and cold dark matter wich is responsible of the formation of structure in the Universe. Observations indicate that stars and dust (baryons) represent something like $0.3\%$ of the whole matter of the Universe. The new measurements of the neutrino mass indicate that neutrinos contribute with a same quantity like matter. In other words, say $\Omega _{M}=\Omega _{m}+\Omega _{DM}=\Omega _{b}+\Omega_{\nu }+\cdot \cdot \cdot +\Omega _{DM}\sim 0.05+\Omega _{DM}$, where $\Omega _{CDM}$ represents the dark matter part of the matter contributions which has a value of $\Omega _{CDM}\sim 0.25$. The value of the amount of baryonic matter ($\sim 5\%$) is in concordance with the limits imposed by nucleosynthesis (see for example \cite{schram}). Then, this model considers a flat Universe ($\Omega_\Lambda + \Omega_M \approx 1$) full with $95\%$ of unknown matter but wich is of great importance at cosmological level. Moreover, it seems to be the most succesful model fitting current cosmological observations\cite{triangle}.

However, from the theoretical point of view, $\Lambda $ has some problems. First, the initial conditions have to be set precisely at one part in $10^{120}$, that is, an extremely fine tuning problem appears. Second, the cosmic coincidence: why is the current value of the energy density contribution of the cosmological constant of the same order than matter?. Third, particle theory predicts a zero cosmological constant, why is it not zero?. These problems can be ameliorated by Quintessence, the model of a fluctuating, inhomogeneous scalar field ($Q$) rolling down a scalar potential $V(Q)$\cite{stein}. It is assumed that flat models with $\Omega _{M}=0.33\pm 0.05$ and a current value of the equation of state $\omega _{Q}=-0.65\pm 0.07$ ($\omega_Q$ can change along the evolution of the Universe) are the most consistent with all observations\cite{quint}. However, there is not agreement about which scalar potential $V(Q)$ is the correct one. For instance, the pure exponential potential has been extensively analyzed\cite{peebles,chimen,ferr,urena,barr,picci}. It is known that there is a solution which makes the scalar energy density scales as the dominant background one, that could help to ameriolate the fine tuning problem. Also, there is another solution that could make the Universe inflate, in good accord with SNIa observations. Moreover, in a scalar dominated Universe, the scalar potential is effectively an exponential one\cite{urena}. But, nucleosynthesis constraints require $\Omega _{Q}\leq 0.2$, and then an exponential potential would never dominate the Universe\cite{ferr}.

Another example is a special group of scalar potentials, named tracker solutions\cite{stein}. The cosmology for these potentials is the same and independent of a large range of initial conditions (about 100 orders of magnitude), avoiding fine tuning. The equation of state $\omega _{Q}$ changes with time towards $-1$\cite{stein,quint}, then it can dominate the evolution of the Universe at late times. A typical example is the pure inverse power-law potential, $V(Q)\sim Q^{-\alpha }$ ($\alpha >0$)\cite{stein,peebles,silviu}. But the predicted value for the current equation of state of the quintessence cannot be put in good accord with supernovae results\cite{stein}. The same problem arises with other inverse power-law-like potentials. Another possibility are the potentials proposed in\cite{varun}. They can solve the troubles stated above, but it is difficult to determine uniquely their free parameters.

On the other side, we do not know the nature of the dark matter component $\Omega _{CDM}$. The $\Lambda $CDM model agrees with the observations of large scale structure of the Universe, but cold dark matter over predicts subgalactic structure and singular cores for the halos of galaxies\cite{galaxs,kamion,vlad}. A scalar field model for dark matter has created a great expectation for solving the problem of the nature of dark matter too\cite{varun,siddh,peebles2,jeremy,hu,luis2}. A scalar field not only could give the correct energy density for the required matter in galaxies to predict the rotation curves of stars, but it is obtained the correct distribution of dark matter in galaxies as well. Intriguinly, a natural solution is an exponential scalar potential\cite{siddh}. At the cosmological level, all attention has been put on the quadratic potential $\Phi^{2}$, because of the well known fact that it behaves as pressureless matter due to its oscillations around the minimum of the potential\cite{ford}, implying that $\omega _{\Phi}\simeq 0$, for $<p_{\Phi }>=\omega _{\Phi }<\rho _{\Phi }>$, just like cold dark matter.

In a recent paper\cite{luis}, we showed that the potential

\begin{eqnarray}
\tilde{V}(\Psi ) &=&\tilde{V_{o}}\left[ \sinh {(\alpha \,\sqrt{\kappa _{o}} \Psi )}\right] ^{\beta }  \label{sinh} \\
&=&\left\{ 
\begin{array}{cc}
\tilde{V_{o}}\left( \alpha \,\sqrt{\kappa _{o}} \Psi \right) ^{\beta } &  |\alpha \,\sqrt{\kappa _{o}} \Psi |\ll 1 \\ 
\left( \tilde{V_{o}}/2^{\beta }\right) \exp {\left( \alpha \beta \,\sqrt{\kappa _{o}} \Psi \right) } & |\alpha \,\sqrt{\kappa _{o}} \Psi |\gg 1
\end{array}
\right. ;  \nonumber
\end{eqnarray}

\noindent is a reliable model for the dark energy, because of its asymptotic behaviors. At the other hand, a good model for dark matter could be the potential\cite{varun,luis2}

\begin{eqnarray}
V(\Phi ) &=&V_{o}\left[ \cosh {(\lambda \,\sqrt{\kappa _{o}} \Phi )}-1\right] \label{cosh} \\
&=&\left\{ 
\begin{array}{cc}
\frac{1}{2} m^2_\Phi \Phi^2 + \frac{1}{24} \lambda^2 m^2_\Phi \kappa_0 \Phi^4 & |\lambda \,\sqrt{\kappa _{o}} \Phi |\ll 1 \\ 
\left( V_{o}/2\right) \exp {\left( \lambda \,\sqrt{\kappa _{o}} \Phi \right)} & |\lambda \,\sqrt{\kappa _{o}} \Phi |\gg 1
\end{array}
\right. .  \nonumber
\end{eqnarray}

\noindent The mass of the scalar field $\Phi$ is defined as $m_{\Phi}^{2}=V^{\prime \prime }|_{\Phi =0}=\lambda ^{2}\kappa _{o}V_{o}$. In this case, we deal with a massive scalar field.

In this paper, we give explicitly all the solutions to the model at the cosmological scale and focus our attention on the scalar dark matter. In section II we find the complete solutions for a cosmology where the scalar field $\Phi$ is the dark matter and the scalar field $\Psi$ is the dark energy. We divide the evolution of the Universe in four stages: the radiation dominated era (RD) but before the oscillations of the scalar field $\Phi$, the radiation dominated era with $\Phi$ already acting as standard cold dark matter, the matter dominated era (MD) and finally the dark energy dominated era. We will recover the standard cosmological evolution of cold dark matter with the quintessence potential (\ref{sinh}). In section III we will analyze the growing fluctuations of the scalar dark matter within the linear regime for the four stages of the cosmological evolution. From the equation of scalar fluctuations, we will determine a Jeans lenght for scalar dark matter and show that modes larger than this Jeans lenght are growing modes. This last fact implies a cut-off in the Mass Power Spectrum and we can fix all free parameters of the potential (\ref{cosh}) by setting the cut-off wave number at the desirable value suggested in the literature. The dark energy fluctuations are not analyzed in detail because it has been already done in the literature. Finally, in section IV we summarize the results and give some future features to be investigated. For completeness, we will use previous results already shown in papers\cite{luis2,luis} when necessary.

\section{Scalar Field Solutions}

Due to current observations of CMBR anisotropy by {\small BOOMERANG} and {\small MAXIMA}\cite{boom}, we will consider a flat, homogenous
and isotropic Universe. Thus we use the flat Friedmann-Robertson-Walker
(FRW) metric

\begin{equation}
ds^2=-dt^2+a^2(t)\left[ dr^2 + r^2 \left(d\theta^2 + \sin^2{(\phi)}d\phi^2
\right) \right] ,
\end{equation}

\noindent where $a$ is the scale factor ($a=1$ today) and we have set $c=1$. The components of the Universe are baryons, radiation, three species of light neutrinos, etc., and two minimally coupled and homogenous scalar fields $\Phi $ and $\Psi $, which represent the dark matter and the dark energy, respectively. The evolution equations for this Universe are

\begin{eqnarray}
H^{2} \equiv \left( \frac{\dot{a}}{a} \right)^2 &=&\frac{\kappa _{o}}{3}
\left( \rho +\rho _{\Phi }+\rho _{\Psi }\right)  \label{fried} \\
\ddot{\Phi}+3H\dot{\Phi}+\frac{dV(\Phi )}{d\Phi } &=&0  \label{cphi} \\
\ddot{\Psi}+3H\dot{\Psi}+\frac{d\tilde{V}(\Psi )}{d\Psi } &=&0
\label{cpsi} \\
{\dot{\rho}}+3H\left( \rho +p\right) &=&0,  \label{fluid}
\end{eqnarray}

\noindent being $\kappa _{o} \equiv 8\pi G$ and $\rho$ ($p$) is the energy density (pressure) of radiation, plus baryons, plus neutrinos, etc. The scalar energy densities (pressures) are $\rho _{\Phi }=\frac{1}{2}\dot{\Phi}^{2}+V(\Phi )$ ($p_\Phi =\frac{1}{2}\dot{\Phi}^{2}-V(\Phi )$) and $\rho_{\Psi }=\frac{1}{2}\dot{\Psi}^{2}+\tilde{V}(\Psi )$ ($p_\Psi =\frac{1}{2}\dot{\Psi}^{2}-\tilde{V}(\Psi )$). Here dots denote derivative with respect the cosmological time $t$.

The numerical values that will be used along the paper are: the Hubble parameter, $H_{o}=100 \, h \, km \, s^{-1} \, Mpc^{-1}$ ($H_o = 3.3 \times 10^{-4} \, h \, Mpc^{-1}$ in units of $c=1$) with $h=0.65\pm0.1$, the current radiation energy density $\Omega_{o\gamma}h^{2}=2.480\times 10^{-5}$, the current baryon energy density $\Omega _{oB}h^{2}=0.019\pm 0.0024$, the background temperature $T_{o}=2.7277\pm 0.002$ K and the amount of primordial helium $Y_{He}=0.246\pm 0.0014$\cite{turner}. We will consider 3 species of light neutrinos with $\Omega_{o\nu}= \frac{7}{8} (4/11)^{4/3} \Omega_{o\gamma}$ per species.

\subsection{Radiation Dominated Era (RD)}

We start the evolution of the Universe at the end of inflation, $i.e.$ in the radiation dominated era (RD). The initial conditions are set such that $\left( \rho_{i\Phi },\,\rho_{i\Psi }\right) \leq \rho_{i\gamma }$.

Let us begin with the dark energy. For the potential (\ref{sinh}) an exact solution in the presence of nonrelativistic matter can be found\cite{chimen,luis}. The parameters of the potential are given by

\begin{eqnarray}
\alpha &=&\frac{-3\omega _{\Psi }}{2\sqrt{3(1+\omega _{\Psi })}},  \nonumber
\\
\beta &=&\frac{2\,(1+\omega _{\Psi })}{\omega _{\Psi }},  \label{tracpsi} \\
\kappa_0 \tilde{V_0} &=& \frac{3(1-\omega_\Psi)}{2} \left( \frac{\Omega_{oM}}{\Omega_{o\Psi}} \right)^{\frac{1+\omega_\Psi}{\omega_\Psi}} \Omega_{o\Psi} H^2_0  \nonumber
\end{eqnarray}

\noindent with $\Omega _{o\Psi }$ and $\Omega _{oCDM}$ the current values of dark energy and dark matter, respectively; and $-0.6\geq \omega _{\Psi }\geq -0.9$ the range for the current equation of state. With these values for the parameters ($\alpha ,\,\beta <0$) the solution for the dark energy ($\Psi $) becomes a tracker one, is only reached until a matter dominated epoch and the scalar field $\Psi $ would begin to dominate the expansion of the Universe after matter domination. Before this, at the radiation dominated epoch, the scalar energy density $\rho _{\Psi }$ is frozen, strongly subdominant and of the same order than today\cite{luis}. Then the dark energy contribution can be neglected during this epoch.

Now we study the behavior of the dark matter. For the potential (\ref{cosh}) we begin the evolution with large and negative values of $\Phi $, when the potential behaves as an exponential one. It is found that the exponential potential makes the scalar field $\Phi$ mimic the dominant energy density, that is, $\rho_\Phi = \rho_{i\Phi} a^{-4}$. The ratio of $\rho_\Phi$ to the total energy density is\cite{chimen,ferr}

\begin{equation}
\frac{\rho _{\Phi }}{\rho _{\gamma }+\rho _{\Phi }}=\frac{4}{\lambda ^{2}}.
\label{rad}
\end{equation}

\noindent This solution is self-adjusting and it helps to avoid the fine tuning problem of matter, too. Here appears one restriction due to nucleosynthesis\cite{ferr}

\begin{equation}
\frac{\rho_\Phi}{\rho_\gamma} = \frac{4}{\lambda^2 -4} < 0.2  \label{lambda}
\end{equation}

\noindent acting on the parameter $\lambda> \sqrt{24}$. We shall call $a^{\ast }$ the scale factor at the time when $|\lambda \,\sqrt{\kappa _{o}}\Phi |\approx 1$, i.e., when the scalar potential (\ref{cosh}) is leaving the exponential behavior and entering into the polynomial one. The time evolution for RD before the scalar oscillations begin is (from eq. (\ref{fried}))

\begin{equation}
H_o t = \frac{\sqrt{\lambda^2-4}}{\lambda} \frac{a^2}{2 \sqrt{\Omega_{o\gamma}}} \, \, (a \leq a^{\ast})
\end{equation}

\noindent where $H_{o}$ is the current value of the Hubble parameter. This result is the standard with an extra contribution due to the scalar contribution $\rho_\Phi$ to radiation.

Once the potential (\ref{cosh}) reaches its polynomial behavior, $\Phi$ oscillates so fast around the minimum of the potential that the Universe is able only to feel the average values of the energy density and pressure in a scalar oscillation. During the time of a scalar oscillation, the term of the cosmological expansion in eq. (\ref{cphi}) can be neglected, then we can write

\begin{equation}
\ddot{\Phi} = - \frac{d V}{d \Phi}.
\end{equation}

\noindent Following\cite{ford}, from the time average in a period larger than a scalar oscillation but smaller than the Hubble time of the quantity

\begin{equation}
\frac{d \left( \Phi \dot{\Phi} \right)}{d t} = \dot{\Phi}^2 + \Phi \ddot{\Phi}
\end{equation}

\noindent it follows that

\begin{equation}
< \dot{\Phi}^2 > = < \Phi \frac{d V}{d \Phi} > .  \label{average}
\end{equation}

\noindent Taking only the quartic and quadratic term, as a good approximation, the average energy density and pressure can be expanded as\cite{peebles2}

\begin{eqnarray}
<\rho _{\Phi }> &\simeq &m_{\Phi }^{2}<\Phi ^{2}>+\frac{1}{8}\lambda^{2}m_{\Phi }^{2}\kappa _{o}<\Phi ^{4}>  \label{apress} \\
<p_{\Phi }> &\simeq &\frac{1}{24}\lambda ^{2}m_{\Phi }^{2}\kappa _{o}<\Phi^{4}>,  \label{arho}
\end{eqnarray}

\noindent then, the equation of state reads

\begin{equation}
<\omega_\Phi> = \frac{1}{3} \frac{<\Phi^4>}{\frac{8}{\lambda^2 \kappa_0}
<\Phi^2> + <\Phi^4>} .  \label{eqstphi}
\end{equation}

\noindent Observe that at the beginning of scalar oscillations, when the quartic term is still the dominant one in the potential, the scalar field $\Phi$ behaves as radiation like the early exponential behavior indicates. Once the quadratic potential dominates, we find that $<\Phi^4> = (3/2) <\Phi^2>^2$ and $<\rho_\Phi> = 2 <V>$. The equation of state then changes to

\begin{equation}
<\omega_\Phi> = \frac{1}{16 V_0} <\rho_\Phi>,  \label{eqst}
\end{equation}

\noindent with $<\omega_\Phi>$ going down to zero and $<\rho_\Phi>$ scaling as non-relativistic matter\cite{ford}. A detailed analysis of the solution of the polynomial behavior can be seen also in\cite{picci}. Now, we would like the scalar field $\Phi $ to act as cold dark matter, in order to recover all the successful features of the standard model. For this to be possible, we will first derive a relation between the parameters ($V_0$, $\lambda$). 

If the transition from radiation to matter occurs smoothly (as suggested by eq. \ref{eqstphi})) at $a=a^\ast$, then

\begin{equation}
\rho_{i\Phi} \left( a^{\ast} \right)^{-4} \approx \rho_{oCDM} \left( a^{\ast} \right)^{-3} \label{limit}.
\end{equation}

\noindent Adjusting numerically and using eq. (\ref{rad}) in the l.h.s of eq. (\ref{limit}), we find that 

\begin{equation}
a^{\ast }\approx \sqrt[3]{\frac{9}{1.7}} \left( \lambda ^{2}-4 \right)^{-1} \left( \frac{\Omega _{o\gamma }}{\Omega _{oCDM}}\right) .  \label{ast}
\end{equation}

\noindent On the other hand, now we use $<\rho_\Phi> \approx \rho_{CDM}$ in the r.h.s of eq. (\ref{limit}). Thus,

\begin{equation}
V_0 <\lambda^2 \kappa_0 \Phi^2>|_{\, a^{\ast}} \approx \rho_{oCDM} \left( a^{\ast} \right)^{-3}.
\end{equation}

\noindent Therefore, the required relation reads\cite{luis2}

\begin{equation}
\kappa_0 V_0 \simeq \frac{1.7}{3} \left( \lambda^2 -4\right)^3 \left(\frac{\Omega_{0CDM}}{\Omega_{o\gamma}} \right)^3 \Omega_{0CDM} H^2_o. \label{ratio}
\end{equation}

\noindent Notice that $V_0$ depends on both current amounts of dark matter and radiation (including light neutrinos) and that we can choose $\lambda$ to be the only free parameter of potential (\ref{cosh}). If we take $\lambda \geq 5$, then we find the limit values

\begin{eqnarray}
\kappa_0 V_0 &\geq& 10^6 \, Mpc^{-2} \\
m_\Phi &\geq& 5 \times 10^3 \, Mpc^{-1} \approx 3 \times 10^{-26} \, eV \\
\lambda \lambda_C &=& (\kappa_0 V_0)^{-1/2} \leq 1 \, kpc \\
< \omega_\Phi > &\sim& 10^{-14} a^{-3} \rightarrow 0,
\end{eqnarray}

\noindent where $\lambda _{C}=m_{\Phi }^{-1}$ is the scalar Compton length. Since now, we can be sure that $\rho_{\Phi}=\rho_{CDM}$ and that we will recover the standard cold dark matter evolution. Moreover, it turns out that the scalar field $\Phi$ is an ultra-light cold dark matter particle. 

The dominant components of the Universe just after scalar oscillations begin are matter and radiation just like in the standard model. Then, we can give the time evolution with radiation and matter only for $a > a^{\ast}$, obtaining

\begin{eqnarray}
H_o \, t = \frac{2}{3} \frac{a^{(3/2)}}{\sqrt{\Omega_{oM}}} \sqrt{1+\frac{a_{\gamma M}}{a}} \left( \frac{1}{2} - \frac{a_{\gamma M}}{a} \right) +  \nonumber \\
\frac{(a^{\ast})^{(3/2)}}{3 \sqrt{\Omega_{oM}}} \sqrt{1+\frac{a_{\gamma M}}{a^{\ast}}} \left( 2 \frac{a_{\gamma M}}{a^{\ast}} -1 \right) + \frac{\sqrt{\lambda^2 -4}}{\lambda} \frac{(a^{\ast})^2}{\sqrt{2 \Omega_{o\gamma}}}, \label{timemd}
\end{eqnarray}

\noindent where

\begin{equation}
a_{\gamma M} = \frac{\Omega_{o\gamma}}{\Omega_{oM}}
\end{equation}

\noindent is the scale factor at the time of radiation-matter equivalence, $a_{\gamma M} \approx 3.3 \times 10^{-4}$ ($z \approx 3000$). The standard case is recovered if $a^{\ast} \rightarrow 0$ ($\lambda \rightarrow \infty$) and that $a^\ast < a_{\gamma M}$ for $\lambda \geq 5$, that is, the transition occurs before radiation-matter equality.

\subsection{Matter Dominated Era (MD)}

During this time, the scalar field $\Phi$ continues oscillating and behaving as nonrelativistic matter and there is a matter dominated era just like that of the standard model. A short after matter completely dominates the evolution of the Universe, the scalar field $\Psi$ reaches its tracker solution, eqs. (\ref{tracpsi}) and\cite{luis}

\begin{eqnarray}
\alpha \,\sqrt{\kappa _{o}} \Psi &=& {\rm arccoth} \left[ 1+\left( \frac{a}{a_{M\Psi }} \right)^{3\omega_\Psi} \right]^{1/2}, \label{phi} \\
\rho_{\Psi} &=& \rho_{o\Psi} a^{-3(1+\omega_{\Psi})},  \label{solmd} \\
\sqrt{\kappa_0} \tilde{V}^{\prime } &=& \alpha \,\beta \,\tanh ^{-1}{\left( \alpha \,\sqrt{\kappa _{o}} \Psi \right) }\,\tilde{V},  \label{first} \\
\sqrt{\kappa_0} \tilde{V}^{\prime \prime } &=& \left[ (\beta -1)\,\tanh ^{-2}{\left( \alpha \, \sqrt{\kappa _{o}} \Psi \right) }+1\right] \alpha ^{2}\,\beta \,\tilde{V}.  \label{second}
\end{eqnarray}

\noindent and it begins to be an important component. $a_{M\Psi}$ is the scale factor at the time of equivalence between matter and dark energy,

\begin{equation}
a_{M \Psi} = \left( \frac{\Omega_{oM}}{\Omega_{o \Psi}} \right)^{-1/(3\omega_\Psi)},
\end{equation}

\noindent where $\Omega _{oM}$ and $\Omega _{o\Psi }$ are the current amounts of matter and dark energy, respectively. We find that $0.62\,(z\approx 0.6)\geq a_{M\Psi }\geq 0.73\,(z\approx 0.37)$ for $-0.6\geq \omega _{\Psi }\geq -0.9$. 

A tracker solution is recognized because the function $\Gamma = (\tilde{V}  \tilde{V}^{\prime \prime})/(\tilde{V}^\prime)^2 > 1$ and nearly constant over the possible initial conditions\cite{stein}. From eqs.~(\ref{first},\ref{second}),

\begin{equation}
\Gamma = \left[ 1 - \beta^{-1}+\beta^{-1} \tanh^2{(\alpha \, \sqrt{\kappa_0} \Psi)} \right],
\end{equation}

\noindent thus $\Gamma \geq 2$ if $\omega_\Psi \leq -0.6$ for the possible initial conditions $0 < \Psi < M_P$. Therefore, potential~(\ref{sinh}) has a tracker solution. The time at wich the scalar field $\Psi$ depends upon initial conditions, but the late time behavior is independent of initial conditions over almost a range of 100 orders of magnitude\cite{stein,luis}.

We can integrate eq. (\ref{fried}) and then the time evolution for matter and dark energy only (neglecting radiation) reads\cite{chimen,luis}

\begin{eqnarray}
H_o \, t = \frac{2}{3(1+\omega_\Psi)} \, \frac{a^{\frac{3}{2} (1+\omega_\Psi)}}{\sqrt{\Omega_{o\Psi}}} \times  \nonumber \\
_2F_1\left( \frac{1}{2},\frac{\beta}{4},\frac{\beta +4}{4}; - \left( \frac{a}{a_{M\Psi}} \right)^{3 \omega_\Psi} \right).  \label{timeone}
\end{eqnarray}

\noindent where ${}_2 F_{1}$ is the hypergeometric function. By analytical continuation, it can be shown that\cite{morse}

\begin{equation}
{}_2 F_1 \left( u,v,w;z \right) = (1-z)^{-u} {}_2 F_1 \left( u,w-v,w; \frac{z}{z-1} \right).
\end{equation}

\noindent If $a \ll a_{M\Psi}$ in eq. (\ref{timeone}), then

\begin{equation}
H_o \, t \rightarrow \frac{2}{3(1+\omega_\Psi)} \frac{a^{(3/2)}}{\sqrt{\Omega_{oM}}} {}_2 F_1 \left( \frac{1}{2}, 1, \frac{\beta}{4} +1; 1 \right).
\end{equation}

\noindent Using eqs. (\ref{tracpsi}), it follows that

\begin{equation}
{}_2 F_1 \left( \frac{1}{2}, 1, \frac{\beta}{4} +1; 1 \right) = 1+\omega_\Psi,
\end{equation}

\noindent therefore

\begin{equation}
H_o \, t \rightarrow \frac{2}{3} \frac{a^{3/2}}{\sqrt{\Omega_{oM}}}. \label{limite}
\end{equation}

\noindent Let us try to give a complete solution for the time in the presence of radiation, matter and dark energy. Taking into account the limit value (\ref{limite}), we can stick up the solutions (\ref{timemd}) and (\ref{timeone}). Then, the complete time evolution for $a > a^{\ast}$ (including the $\Psi$ dominated era) can be written as

\begin{eqnarray}
H_o \, t = \frac{2}{3(1+\omega_\Psi)} \, \frac{a^{\frac{3}{2} (1+\omega_\Psi)}}{\sqrt{\Omega_{o\Psi}}} \left[ 1+ \left( \frac{a}{a_{M\Psi}} \right)^{3 \omega_\Psi} \right]^{-1/2} \times  \nonumber \\
{}_2F_1\left( \frac{1}{2},1,\frac{\beta +4}{4}; \frac{(a/a_{M\Psi})^{3\omega_\Psi}}{1+(a/a_{M\Psi})^{3 \omega_\Psi}} \right) \sqrt{1+\frac{a_{\gamma M}}{a}} \left( \frac{1}{2} - \frac{a_{\gamma M}}{a} \right) \nonumber \\
- \frac{2}{3} \frac{(a^{\ast})^{3/2}}{\sqrt{\Omega_{oM}}} \sqrt{1+\frac{a_{\gamma M}}{a^{\ast}}} \left( \frac{1}{2} - \frac{a_{\gamma M}}{a^{\ast}} \right) + \frac{\sqrt{\lambda^2 -4}}{\lambda} \frac{(a^{\ast})^2}{2 \sqrt{\Omega_{o\gamma}}}.  \label{age}
\end{eqnarray}

\subsection{Scalar Field $\Psi$ Dominated Era ($\Psi$D)}

At this time, the scalar field $\Psi$ is the dominant componente of the Universe and the scalar potential (\ref{sinh}) is effectively an exponential one\cite{urena,luis}. The time evolution becomes 

\begin{equation}
H_o \, t = \frac{2}{3(1+\omega_{\Phi})} \, \frac{ a^{\frac{3}{2} (1+\omega_{\Phi})}}{\sqrt{\Omega_{o\Phi}}},  \label{SFD5}
\end{equation}

\noindent thus the scalar field $\Psi$ drives the Universe into a power-law inflationary stage ($a \sim t^p$, $p>1$). Note that usual tracker solutions do not have this late exponential behavior. Also, tracker equation of state usually changes toward $-1$ once the scalar field dominates finishing the latter as a cosmological constant\cite{stein}. But the scalar equation of state $\omega_\Psi$ will never change its tracker value and for potential (\ref{sinh}) there is no solution with $\omega_\Psi =-1$ either\cite{luis}.

\begin{figure}[h]
\centerline{ \epsfysize=5cm \epsfbox{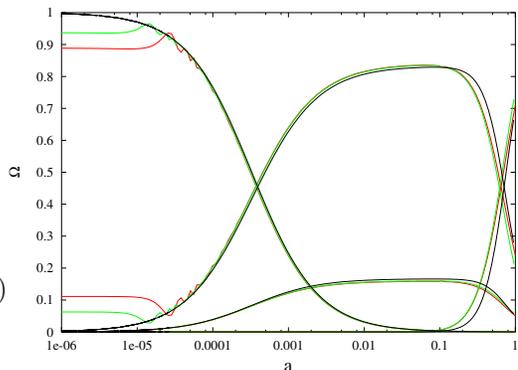}}
\caption{Evolution of the dimensionless density parameters {\it vs} the scale factor $a$ with $\Omega_{oM} = 0.30$: $\Lambda$CDM (black) and $\Psi\Phi DM$ for two values of $\protect\lambda=6$ (red), $\protect\lambda=8$ (green). The equation of state for the dark energy is $\protect\omega_\Psi = -0.8$.} 
\label{fig:Omegac}
\end{figure}

A complete numerical solution for the energy densities $\rho$'s and the dimensionless density parameters $\Omega $'s are shown in fig.~\ref{fig:Omegac} until today. The results agree with the solutions found in this section. It can be seen that eq. (\ref{ratio}) makes the scalar field $\Phi$ behave quite similar to the standard cold dark matter model once the scalar oscillations begin and the required contributions of dark matter and dark energy are the observed ones\cite{luis2,luis}.

\begin{figure}[h]
\centerline{ \epsfysize=5cm \epsfbox{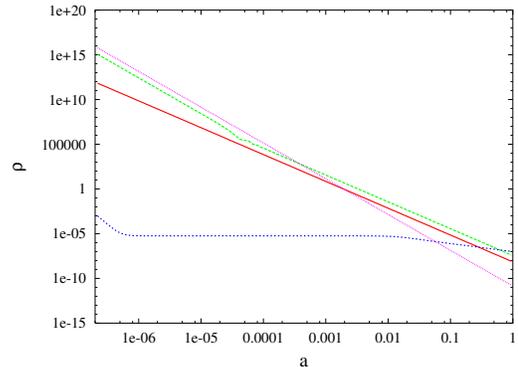}}
\caption{Evolution of the energy densities $\rho_i$ {\it vs} $a$: $\protect\rho_\protect\protect\gamma$ (pink), $\protect\rho_{CDM}$ (green), $\protect\rho_b$ (red) and $\protect\rho_\Psi$ (blue); with $\protect\omega_\Psi = -0.7$.}
\label{fig:rhos}
\end{figure}

\noindent The current age of the Universe can be calculated from eq. (\ref{age}) with $a=1$. Thus, $1.09 \geq H_o \, t_o \geq 4.02$ for $-0.6 \geq \omega_\Psi \geq -0.9$. The age of the Universe could be $(15 \leq t_o \leq 60) \times 10^9$ years, irrespectively of the value of $\lambda$. We recovered the standard cosmological evolution and then we can see that potentials (\ref{sinh},\ref{cosh}) are reliable models of dark energy and dark matter in the Universe.

\section{Linear Perturbation Theory}

In this section, we analyze the perturbations of the space due to the presence of the scalar fields $\Phi,\Psi$ using analytical approximations. First, we consider a linear perturbation of the space given by $h_{ij}$. We will work in the synchronous gauge formalism, where the line element is $ds^{2}=a^{2}[-d\tau ^{2}+(\delta_{ij}+h_{ij})dx^{i}dx^{j}]$. The equations for perturbations in the $k$-space ($\vec{k} = k \hat{k}$) are\cite{ma}

\begin{eqnarray}
h_{ij}(\vec{x},\tau)=  \nonumber \\
\int d^3 k e^{i\vec{k} \cdot \vec{x}} \left\{ \hat{k_i} \hat{k_j} h\left( \vec{k},\tau \right) + \left( \hat{k_i} \hat{k_j} - \frac{1}{3} \delta_{ij} \right) 6 \eta(\vec{k},\tau) \right\},
\end{eqnarray}

\begin{eqnarray}
k^2 \eta - \frac{1}{2} \frac{\dot{a}}{a} \dot{h} &=& 4 \pi G a^2 \delta T^0_0, \label{evol1} \\
k^2 \eta &=& 4 \pi G a^2 (\rho + p) \theta,  \label{evol2} \\
\ddot{h} + 2 \frac{\dot{a}}{a} \dot{h} - 2 k^2 \eta &=& -8 \pi a^2 \delta T^{i}_{i},  \label{evol3} \\
\ddot{h} + 6 \ddot{\eta} + 2 \frac{\dot{a}}{a} \left( \dot{h} + 6 \dot{\eta} \right) - 2 k^2 \eta &=& -24 \pi G a^2 (\rho + p) \sigma ;
\end{eqnarray}

\noindent where $h$ is the trace of the metric perturbations $h_{ij}$; $\delta T^\mu_\nu$ is the perturbation to the momentum-energy tensor and now dots denote derivatives with respect to conformal time $\tau$. The velocity $\theta$ and shear $\sigma$ perturbations are defined as

\begin{eqnarray}
(\rho + p) \theta &\equiv & i k^j \delta T^0_j, \\
(\rho + p) \sigma &\equiv & - \left( k_i k_j - \frac{1}{3} \delta_{ij} \right) \Sigma^i_j, \\
\Sigma^i_j &\equiv & T^i_j - \frac{1}{3} \delta^i_j T^k_k .
\end{eqnarray}

\noindent The density contrast ($\delta \equiv \delta \rho / \rho$) wich accounts for the over density relative to the homogeneus cosmological energy density and $\theta$ for perfect fluids are governed by

\begin{eqnarray}
\dot{\delta} &=&-\left( 1+\omega \right) \left( \theta +\frac{\dot{h}}{2} \right) -3\frac{\dot{a}}{a}\left( \frac{\delta p}{\delta \rho }-\omega \right) \delta,  \label{delta} \\
\dot{\theta} &=&-\frac{\dot{a}}{a}\left( 1-3\omega \right) \theta -\frac{\dot{\omega}}{1+\omega }\theta +\frac{\delta p/\delta \rho }{1+\omega } k^{2}\delta -k^{2}\sigma .
\end{eqnarray}

\noindent The evolution equations with the unperturbed FRW metric (\ref{fried}-\ref{fluid}) in terms of the conformal time are

\begin{eqnarray}
{\cal H}^{2} = \frac{\kappa _{o}}{3}a^{2} \left( \rho + \rho_\Phi +
\rho_\Psi \right) \\
\ddot{\Phi}+2{\cal H}\dot{\Phi}+a^2 \frac{dV(\Phi )}{d\Phi } =0 \\
\ddot{\Psi}+2{\cal H}\dot{\Psi}+a^2 \frac{d\tilde{V}(\Psi )}{d\Psi } =0 \\
{\dot{\rho}}+3{\cal H}\left( \rho +p\right) =0,
\end{eqnarray}

\noindent ${\cal H}$ being the conformal Hubble factor, the scalar energy densities are $\rho_\Phi =(1/2 a^2)\dot{\Phi}^{2}+V(\Phi)$ and $\rho_{\Psi }=(1/2a^2)\dot{\Psi}^{2}+\tilde{V}(\Psi)$. The scalar pressures are $p_\Phi =(1/2 a^2)\dot{\Phi}^{2}-V(\Phi )$ and $p_\Psi =(1/2 a^2)\dot{\Psi}^{2}-\tilde{V}(\Psi )$. The solution to this equations are those already found in the previous section when written as functions of the scale factor $a$

\subsection{Scalar Perturbations}

We must add the perturbed equations for the scalar fields $\Phi(\tau) \rightarrow \Phi(\tau) + \phi(k,\tau)$ and $\Psi(\tau) \rightarrow \Psi(\tau) + \psi(k,\tau)$\cite{stein,ferr}

\begin{eqnarray}
\delta \rho_{\Phi} &=& \frac{1}{a^2} \dot{\Phi} \dot{\phi} + V^\prime \phi, \nonumber \\
\delta p_{\Phi} &=& \frac{1}{a^2} \dot{\Phi} \dot{\phi} - V^\prime \phi, \\
(\rho_{\Phi} + p_{\Phi}) \theta_{\Phi} &=& \frac{1}{a^2} \dot{\Phi} k^2 \phi , \nonumber
\end{eqnarray}

\begin{eqnarray}
\delta \rho_{\Psi} &=& \frac{1}{a^2} \dot{\Psi} \dot{\psi} + \tilde{V}^\prime \psi  \nonumber \\
\delta p_{\Psi} &=& \frac{1}{a^2} \dot{\Psi} \dot{\psi} - \tilde{V}^\prime \psi \\
(\rho_{\Psi} + p_{\Psi}) \theta_{\Psi} &=& \frac{1}{a^2} \dot{\Psi} k^2 \psi , \nonumber
\end{eqnarray}

\noindent with the evolution equations for the perturbations

\begin{eqnarray}
\ddot{\phi} + 2 {\cal H}\dot{\phi} + k^2 \phi + a^2 V^{\prime \prime} \phi + \frac{1}{2} \dot{\Phi} \dot{h} &=& 0  \label{pphi} \\
\ddot{\psi} + 2 {\cal H} \dot{\psi} + k^2 \psi + a^2 \tilde{V}^{\prime \prime} \psi + \frac{1}{2} \dot{\Psi} \dot{h} &=& 0 . \label{ppsi}
\end{eqnarray}

\noindent Here, primes are derivatives with respect of the unperturbed scalar fields $\Phi$ and $\Psi$, respectively. Before we solve these equations, we analyze first the meaning of a scalar Jeans lenght.

\subsection{Damping of the Scalar Power Spectrum}

Recalling the results for the average scalar pressure and energy density (eqs.~\ref{apress},\ref{arho}), the velocity of sound in the scalar fluid $\Phi$ is

\begin{equation}
v^2_s = \frac{1}{8 V_0} <\rho_\Phi> \sim 10^{-14} a^{-3}  \label{vs}
\end{equation}

\noindent the velocity of sound decreases rapidly to zero as $\omega_{\Phi }$ does. As suggested in the literature, we could define an effective Jeans length\cite{varun,peebles2,pad2},

\begin{equation}
L_J = \sqrt{\pi} \sqrt{\frac{v^2_s}{G <\rho>}} = \pi \left(\kappa_0 V_0 \right)^{-1/2} \label{jeans}
\end{equation}

\noindent from which we obtain that $L_{J} \leq 3.14 \, kpc$. Therefore, we can expect that the scalar field fluctuations produce only complex stelar objects bigger that this Jeans length. In first approximation, this model could explain the suppression of subgalactic structure\cite{varun}. Also, it is likely that the quantity $r_{c}=(\kappa_{o}V_{o})^{-1/2}$ could be important at galactic scale\cite{luis2}. But, in order to have a good estimate of the real Jeans lenght, we need to analyze the flutuation equations. We will find that eq. (\ref{jeans}) is only an approximation to the real thing.

It is known that scalar perturbations can only grow if the $k^{2}$-term in eqs. (\ref{pphi},\ref{ppsi}) is subdominant with respect to the second derivative of the scalar potential, that is, if $k<aV^{\prime \prime}$\cite{ma2}. Let us anlyze the scalar dark matter case.

\begin{figure}[h]
\centerline{ \epsfysize=5cm \epsfbox{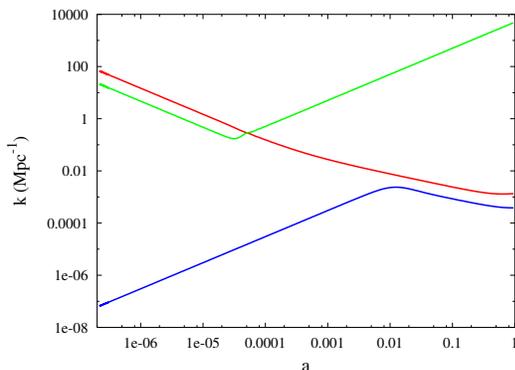}}
\caption{Evolution of the Hubble wave number $k_H$ (red) and the scalar wave numbers $k_\Phi$ (green) and $k_\Psi$ (blue).}
\label{fig:dkas}
\end{figure}

According to the solution given above for potential (\ref{cosh}), the behavior for $k_{\Phi }=aV^{\prime \prime }$ is

\begin{equation}
k_\Phi (a) = \left\{ 
\begin{array}{cc}
2 H_o \sqrt{\frac{\Omega_{o\gamma}}{\lambda^2-4}} \lambda a^{-1} & (a < a^{\ast}) \\ 
m_\Phi a & (a^{\ast} \leq a)
\end{array}
\right.
\end{equation}

\noindent with a minimum value given by

\begin{equation}
k_{min,\Phi} = m_\Phi a^{\ast} \simeq 1.3 \, \lambda \sqrt{\lambda^2 -4} \frac{\Omega_{0CDM}}{\sqrt{\Omega_{o\gamma}}} H_o , \label{kmin}
\end{equation}

\noindent and then for $\lambda \geq 5$, $k_{min,\Phi} \geq 0.375 \, Mpc^{-1}$ (see fig~\ref{fig:dkas}).

\noindent It can be assured that there are no scalar perturbations for $k > m_\Phi$, that is, bigger than $k_\Phi$ today. These $k$ corresponds to scales smaller than $1.2 \, kpc$ (here $\lambda=5$). They must have been completely erased. Besides, modes which $m_\Phi > k > k_{min,\Phi}$ must have been damped during certain periods of time determined by

\begin{equation}
2 H_o \, \lambda \sqrt{\frac{\Omega_{o\gamma}}{\lambda^2-4}} \, k < a < 
\frac{k}{m_\Phi}.
\end{equation}

\noindent From this, we conclude that the scalar power spectrum of $\Phi$ will be damped for $k > k_{min,\Phi}$ with respect to the standard case. Also, in fig.~\ref{fig:dkas} we can see that the Hubble wave number $k_H$ is greater than $k_\Phi$ until a little after $a^{\ast}$. It means the Jeans length (\ref{jl}) is well within the Hubble horizon for most of the time, then the scalar fluctuatinos can form structure at scales smaller than the Hubble lenght $H^{-1}$.

Therefore, the real Jeans length must be

\begin{equation}
L_J (a) = 2 \pi \, k^{-1}_{min,\Phi},  \label{jl}
\end{equation}

\noindent and it is a universal constant because it is completely determined by the mass of the scalar field particle. $L_J$ is not only proportional to the quantity $(\kappa_0 V_0)^{-1/2}$, as suggested by the approximation in eq. (\ref{jeans}), but also the time when scalar oscillations start (represented by $a^\ast$) is important, as it was suspected in\cite{varun}.

On the other hand, the wave number $k_{\Psi }=a\tilde{V}^{\prime \prime }$ is always out of the Hubble horizon, then only structure at larger scales than $H^{-1}$ can be formed by the scalar fluctuations $\psi$. Instead of a minimum, there is a maximum $k_{max,\Psi }\sim 10^{-3}$. All scalar perturbations of the dark energy which $k>10^{-3}$ must have been completely erased. Perturbations with $k\leq 10^{-3}$ have started to grow only recently (see fig.~\ref{fig:dkas}). For a more detailed analysis of the dark energy fluctuations, see\cite{ma2,brax}.

\subsection{Scalar Power Spectrum for $\Phi$}

In this subsection, we will find some analytical approximated solutions to eq. (\ref{pphi}), having in mind the Jeans Length (\ref{jl}), that is, we will analyze eq. (\ref{pphi}) only for modes $k<k_{min,\Phi }$.

We can have some physical insight into the previous differential equations if we write the evolution equations (\ref{evol1},\ref{evol3}) in the form

\begin{eqnarray}
\frac{d}{d \tau} \left( a \dot{h} \right) = \kappa_0 a^3 \left( \delta T^0_0 - \delta T^j_j \right),
\end{eqnarray}

\noindent then we can change this into

\begin{eqnarray}
\frac{d}{d \tau} \left( a \dot{h} \right) +3 a {\cal H}^2 \left[2\Omega_{\gamma} \delta_{\gamma} + \Omega_b \delta_b + \right.  \nonumber \\
\left. \Omega_{\Phi} \delta_{\Phi} \left( 1 + 3 \frac{\delta p_{\Phi}}{\delta \rho_{\Phi}} \right) + \Omega_{\Psi} \delta_{\Psi} \left( 1 + 3 \frac{\delta p_{\Psi}}{\delta \rho_{\Psi}} \right) \right] = 0 ,  \label{hdot}
\end{eqnarray}

\noindent where $\Omega_i = \rho_i/ \rho_T$, $\rho_T = (3 {\cal H}^2)/(\kappa_0 a^2)$. For perfect fluids, it happens that $(\delta p / \delta \rho)= v^2_s \approx \omega$. For scalar fields, we can not identify $v^2_s \approx \omega$ in general and so we have explicitly written $(\delta p_{\Phi,\Psi} / \delta \rho_{\Phi,\Psi})$.

We can see easily that the dominant background component of the Universe is the dominant term in the differential equation (\ref{hdot}), too. Thus, we can use the standard results for the fluctuations during ceratin stages of the cosmological evolution. Also, it is worth to mention that all perturbed quantities can be written in terms of the trace of the metric perturbations $h$, then it is enough to have a solution for it. See \cite{ma,seljak} for the other equations required and how to solve them.

\subsubsection{$a < a^{\ast}$: Radiation Dominated Era}

During RD, the scalar energy density $\rho_\Phi$ evolve as a perfect fluid with constant equations of state $\omega_\Phi = 1/3$ (remember that the exponential potential mimics the dominant energy). For the scalar field $\Phi $ we have

\begin{eqnarray}
\kappa_0 a^2 \rho_\Phi &=& \frac{12}{\lambda^2} {\cal H}^2 \\
\sqrt{\kappa_0} \dot{\Phi} = \sqrt{\frac{4}{3} a^2 \kappa_0 \rho_{\Phi}} &=& \frac{4}{\lambda} {\cal H}  \nonumber \\
a^2 V^{\prime} = - {\cal H} \, \sqrt{\kappa_0} \dot{\Phi} &=& - \frac{4}{\lambda} \frac{{\cal H}^2}{\kappa_0}  \label{utils1} \\
a^2 V^{\prime \prime} &=& 4 {\cal H}^2.  \nonumber
\end{eqnarray}

\noindent Since ${\cal H}=\tau^{-1}$ and radiation dominates equation (\ref{hdot}), $h$ evolves as in the standard case, $\dot{h}=C \tau$ for modes out the Hubble horizon and $\dot{h}=C \tau^{-1}$ for modes inside the Hubble horizon\cite{pad} (for a detailed calculus with the exponential potential, see\cite{ferr}), being $C=const.$. 

Having that $a=\tau$, the evolution equation (\ref{pphi}) for a mode out of the Hubble horizon can be written (see eqs. \ref{utils1})

\begin{equation}
\ddot{\phi} + \frac{2}{\tau}\dot{\phi} + \frac{4}{\tau^2} \phi = -\frac{2C}{\sqrt{\kappa_0} \lambda}.
\end{equation}

\noindent Thus, the growing solution for $\phi$ is

\begin{equation}
\sqrt{\kappa_0} \phi(\tau) = - \frac{C}{5 \lambda} \tau^2 ,
\end{equation}

\noindent where we can recover the result $-(1/2)h=\delta _{CDM}=(5\lambda/4)\,\sqrt{\kappa _{o}}\phi $\cite{ferr}. Here, $\delta _{CDM}$ would be the standard CDM density contrast. This leads to

\begin{equation}
\delta_\Phi = \frac{4}{15} \delta_{CDM} ,  \label{ecdm}
\end{equation}

\noindent that is, the scalar density contrast $\delta_\Phi$ evolves as the standard one but with smaller amplitude. Note that this result is independent of $\lambda$. This result can also be obtained from eq. (\ref{delta}) with

\begin{equation}
\frac{\delta p_\Phi}{\delta \rho_\Phi} = 3, \, \omega_\Phi = \frac{1}{3},
\end{equation}

\noindent where we can observe that the scalar field $\Phi$ does not behave completely as perfect fluid.

For modes inside the Hubble horizon, the equation to be solved is

\begin{equation}
\ddot{\phi} + \frac{2}{\tau}\dot{\phi} + \frac{4}{\tau^2} \phi = -\frac{2C}{\sqrt{\kappa_0} \lambda} \frac{1}{\tau^2}.
\end{equation}

\noindent The general solution is of the form

\begin{equation}
\sqrt{\kappa_0} \phi(\tau) = - \frac{C}{2 \lambda}=const.,
\end{equation}

\noindent and then $\delta_\Phi = (C/6) = const$. The modes inside the Hubble horizon do not grow during RD.

\subsubsection{$a^{\ast} < a$: Radiation and Matter Dominated eras}

Once the scalar field $\Phi$ begins to oscillate, it happens that $V^{\prime \prime} = m^2_\Phi$. Recalling that $k < k_{min,\Phi} < a^2 V^{\prime \prime} $, eq. (\ref{pphi}) can be written

\begin{equation}
\ddot{\phi} + 2 {\cal H} \dot{\phi} + a^2 m^2_\Phi \phi + \frac{1}{2} \dot{\Phi} \dot{h} = 0 .  \label{pphi2}
\end{equation}

\noindent The scalar perturbation $\phi$ oscillates with the same frecuency that the unperturbed $\Phi$. Following the previous section, if we take the time average of the quantity

\begin{equation}
\frac{d (\Phi \dot{\phi})}{d \tau} = \dot{\Phi} \dot{\phi} + \Phi \ddot{\phi}
\end{equation}

\noindent we obtain

\begin{equation}
< \dot{\Phi} \dot{\phi} > = - < \Phi \ddot{\phi} >.
\end{equation}

\noindent The second and fourth terms of eq. (\ref{pphi2}) are almost constant during the time of a scalar oscillation. Then,

\begin{equation}
< \Phi \ddot{\phi} > \approx - < a^2 m^2_\Phi \Phi \phi > = - < a^2 V^{\prime} \phi >
\end{equation}

\noindent Therefore, we find that

\begin{eqnarray}
\frac{<\delta p_\Phi>}{<\delta \rho_\Phi>} = \frac{< \dot{\Phi} \dot{\phi} > - < a^2 V^{\prime} \phi >}{< \dot{\Phi} \dot{\phi}> + < a^2 V^{\prime} \phi > } \approx 0.
\end{eqnarray}

\noindent It is now convenient to rewrite eq. (\ref{pphi2}) in the same form
as eq. (\ref{delta}),

\begin{equation}
\dot{\delta_\Phi} + 3 {\cal H} \delta_\Phi \left( \frac{< \delta p_\Phi >}{< \delta \rho_\Phi >} - \frac{< p_\Phi >}{< \rho_\Phi >} \right) = - \left( 1+\frac{< p_\Phi >}{< \rho_\Phi >} \right) \frac{\dot{h}}{2}
\end{equation}

\noindent Therefore, with $<\omega_\Phi>$ going to zero (eq.~\ref{eqst}) it follows

\begin{equation}
\delta_\Phi = \delta_{CDM} .  \label{scdm}
\end{equation}

\begin{figure}[h]
\centerline{ \epsfysize=5cm \epsfbox{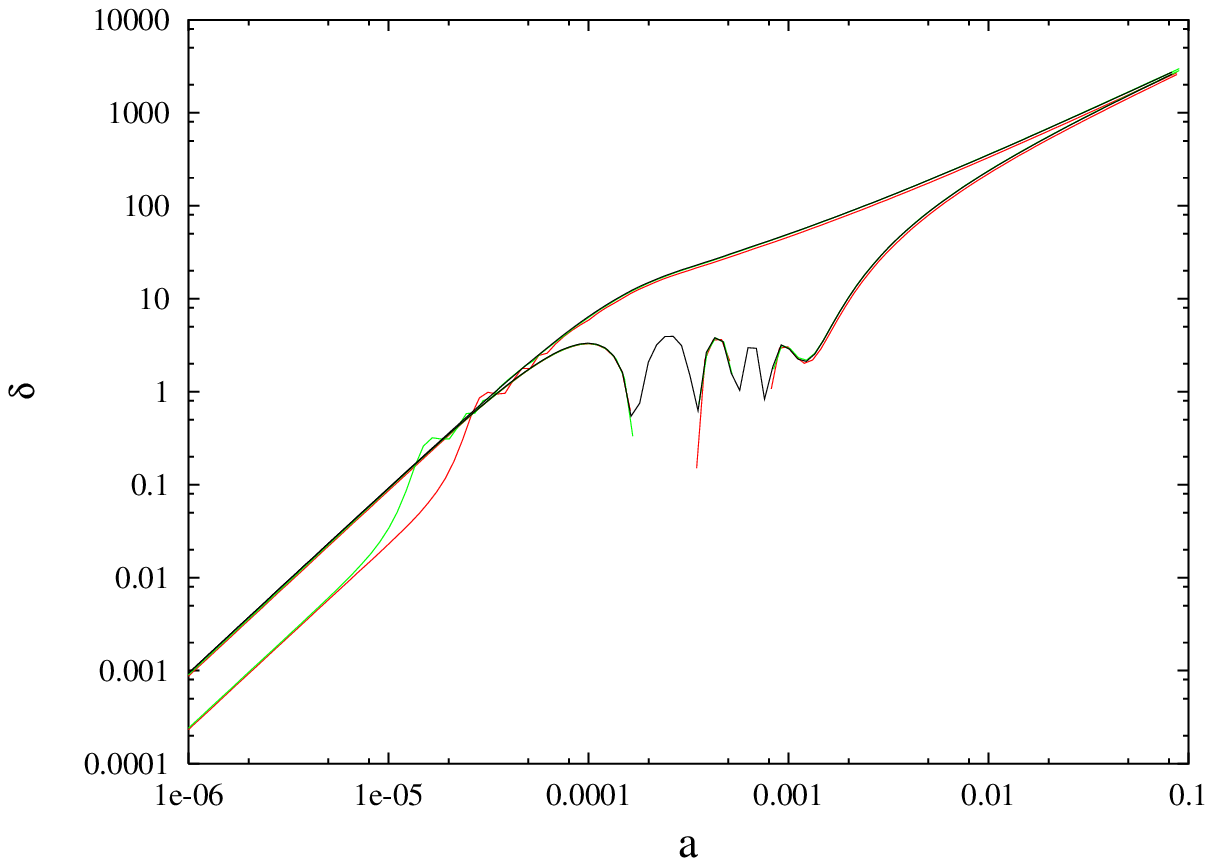}}
\centerline{ \epsfysize=5cm \epsfbox{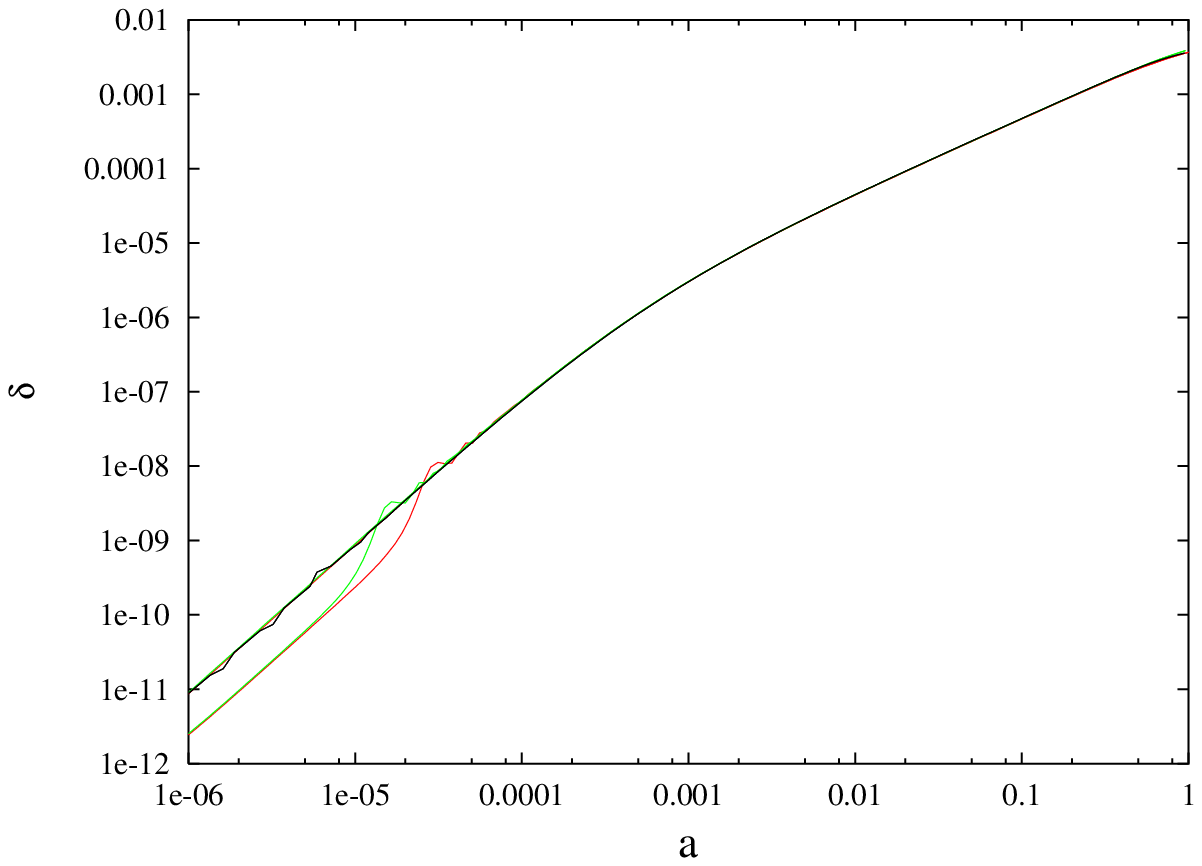}}
\caption{Evolution of the density contrasts for baryons $\delta_b$, standard cold dark matter $\delta_{CDM}$ and scalar dark matter $\delta_\Phi$ {\it vs} the scale factor $a$ taking $\Omega_{oM} = 0.30$ for the models given in fig.~(\ref{fig:Omegac}). The modes shown are $k=0.1 \, Mpc^{-1}$ (top) and $k=1.0 \times 10^{-5} \, Mpc^{-1}$ (bottom).}
\label{fig:dphi}
\end{figure}

\noindent Due to its oscillations around the minimum, the scalar field $\Phi$ changes to a complete standard CDM and so do its perturbations. All the standard growing behavior for modes $k < k_{min,\Phi}$ is recovered and preserved until today by potential (\ref{cosh}). In fig.~\ref{fig:dphi}, a numerical evolution of $\delta \equiv \delta \rho / \rho$ is shown compared with the standard CDM case\cite{luis2}. The results (\ref{ecdm},\ref{scdm}) agree with the numerical solution. The numerical evolution for the density contrasts was done using an amended version of {\small CMBFAST}\cite{seljak}.

\begin{figure}[h]
\centerline{ \epsfysize=5cm \epsfbox{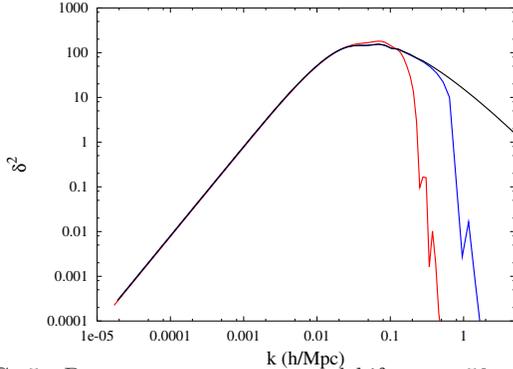}}
\caption{Power spectrum at a redshift $z=50$: $\Lambda$CDM (black), and $\Phi $CDM with $\protect\lambda = 5$ (red) and $\protect\lambda = 10$ (blue). The normalization is arbitrary.}
\label{fig:tf}
\end{figure}

\noindent In fig.~\ref{fig:tf} we can see $\delta ^{2}$ at a redshift $z=50$ from a complete numerical evolution using the amended version of {\small CMBFAST}. We also observe a sharp cut-off in the processed power spectrum at small scales when compared to the standard case, as it was argued above. This suppression could explain the smooth cores of dark halos in galaxies and a less number of dwarf galaxies\cite{kamion}. 

At this point, we would like to mention some coincidences between some results in ref.\cite{hu} in wich a quadratic potential is used and the results of this section. In that reference, it is argued that the Jeans lenght is the de Broglie wavelenght at the ground state of the particle in the gravitational potential well and that the power spectrum is supressed relative to the CDM case. We see before that most of the interesting properties of potential (\ref{cosh}) as dark matter are due to its polynomial behavior $\Phi^2$. Then, it is not strange that, in our case, the mass power spectrum is also related to the CDM case by (see\cite{hu})

\begin{equation}
P_\Phi(k) \simeq \left( \frac{\cos{x^3}}{1+x^8} \right)^2 P_{CDM}(k),
\end{equation}

\noindent but using $x=(k/k_{min,\Phi})$ with $k_{min,\Phi}$ being the wave number associated to the Jeans lenght (\ref{jl}). The difference with respecto to the case treated in\cite{hu} is that the relevant time scale is that when scalar oscillations start and not that of radiation-matter equality.

If we take a cut-off of the mass power spectrum at $k=4.5 \, h \, {\rm{Mpc}}^{-1}$\cite{kamion}, we can fix the value of parameter $\lambda$. Using eq. (\ref{kmin}), we find that 

\begin{eqnarray}
\lambda &\simeq& 20.28, \nonumber \\
V_0 &\simeq& \left(3.0\times 10^{-27}\,M_{Pl}\simeq 36.5\, {\rm{eV}}\right) ^{4}, \\
m_\Phi &\simeq& 9.1\times 10^{-52}\,M_{Pl}\simeq 1.1\times 10^{-23}\, {\rm{eV}}. \nonumber 
\end{eqnarray}

\noindent where $M_{Pl}=1.22 \times 10^{19} \, {\rm{GeV}}$ is the Plank mass. All parameters of potential (\ref{cosh}) are now completely determined and we have the right cut-off in the mass power spectrum.

\subsubsection{Scalar Field $\Psi$ Dominated Era}

Only for completeness, we will draw some general features of the evolution of fluctuations during the dark energy dominated era.

At this era, the scalar field $\Psi$ now dominates both the evolution of the Universe and the differential equation (\ref{hdot}). We do not worry anymore of $\Phi $, because its perturbed solution continues being $\delta _{\Phi}=-(1/2) h$ due to its oscillations around the minimum of the potential. The scalar energy $\rho _{\Psi }$ evolve as a perfect fluid with equation of state $\omega _{\Psi }$ due to the effective exponential behavior of potential (\ref{sinh}). Then,

\begin{eqnarray}
\dot{\Psi} &=& \sqrt{1+\omega_\Psi} \sqrt{a^2 \rho_\Psi} \\
\tilde{V} &=& \frac{1-\omega_\Psi}{2} \rho_\Psi \\
\tilde{V}^{\prime} &=& \sqrt{3(1+\omega_\Psi)} \tilde{V} \\
\tilde{V}^{\prime \prime} &=& 3(1+\omega_\Psi) \tilde{V} .
\end{eqnarray}

\noindent Since the scalar field $\Psi $ dominates the evolution of the Universe, it is straightforward that

\begin{eqnarray}
a^2 \rho_\Psi &=& \frac{3 {\cal H}^2}{\kappa_0} \\
{\cal H} &=& \frac{2}{1+3\omega _{\Phi }}(\tau-\tau _{\infty})^{-1}
\end{eqnarray}

\noindent being $\tau_{\infty}$ the size of the event horizon ($\tau \rightarrow \tau_{\infty}$)\cite{ma}. Since we are interested in possible growing modes, we will consider eq. (\ref{ppsi}) only in the case $k^2 \ll a^2  \tilde{V}^{\prime \prime}$. The evolution equations (\ref{ppsi},~\ref{hdot}) become:

\begin{eqnarray}
\ddot{\psi} + 2 B \frac{\dot{\psi}}{(\tau_{\infty}-\tau)} + C \frac{\psi}{(\tau_{\infty}-\tau)^2} + \frac{D}{2} \frac{\dot{h}}{(\tau_{\infty}-\tau)} &=& 0 \\
\ddot{h} + B \frac{\dot{h}}{(\tau_{\infty}-\tau)} + 4 D \frac{\dot{\psi}}{(\tau_{\infty}-\tau)} - E \frac{\psi}{(\tau_{\infty}-\tau)^2} &=& 0
\end{eqnarray}

\noindent where the constant coefficients are

\begin{eqnarray}
B &=& \frac{2}{1+3\omega _{\Phi }} \\
C &=& \frac{9}{2} (1-\omega^2_\Psi) B^2 \\
D &=& \sqrt{3(1+\omega_\Psi)} B \\
E &=& \frac{3}{2} \sqrt{3(1+\omega_\Psi)} (1-\omega_\Psi) B^2.
\end{eqnarray}

\noindent The scalar fields were normalized in units of $\kappa_0^{-1/2}$. We can try solutions of the form $h=h_o (\tau-\tau_{\infty})^{m}$, $\psi=\psi_o (\tau-\tau_{\infty})^{m}$, where the values of $m$ are the solutions to the equation

\begin{equation}
m \left[ m (m-1) + 2Bm+C \right] \left[ (m-1)+B \right]= \frac{D m}{2} (4Dm-E)
\end{equation}

\noindent with an obvious solution $m=0$ and two complex roots . The other real root gives $m>0$ if $-0.6 \leq \omega_\Psi \leq -0.9$. For instance, if  $\omega_\Psi=-0.6$, $m=\{ 7.83, \, 0.835 + 3.2i, \, 0.835 - 3.2i\}$. Then, all of the solutions are decaying ones. This result is not surprising because by this time the Universe has already entered in an inflationary stage. Then, we conclude that perturbations in the linear regime will be suppressed by the accelerated expansion of the Universe.

\section{Conclusions}

We have developed most of the interesting features of a $95\%$ scalar-nature cosmological model. The interesting implications of such a model are direct consequences of the scalar potentials (\ref{sinh},\ref{cosh}). 

On one hand, we have modeled the dark energy of the Universe using a scalar field with a $\sinh^{\beta }$ potential\cite{luis}. This potential has the advantages that at early times it is a pure inverse power-law one, $\tilde{V}(\Psi )\sim \Psi ^{-\alpha }$, thus the cosmology at late times is extremely insensitive to initial conditions, reducing the fine tuning. The $\Psi $ solution is only reached during a matter dominated phase of the Universe, thus it is natural the existence of a current dark energy dominated epoch; the cosmic coincidence is ameliorated. Its late exponential behavior drives the Universe into a power-law inflationary stage, in good accord with supernovae results. Nevertheless, a fine tunning problem remains in determining the parameters of the potential. Also, we do not know about a fundamental theory that could predict this kind of potential. However, quintessence models with a expectation value of the field of the order of the Planck mass can be considered within supergravity\cite{cope2}.

On the other hand, we have modeled the cosmological dark matter using another scalar field $\Phi $ with a $\cosh $ scalar potential. As we have shown in this work, the solutions found alleviate the fine tuning problem for cold dark matter, too. Once the scalar field $\Phi $ begins to oscillate around the minimum of its potential (\ref{cosh}), we can recover the evolution of a standard cold dark matter model because the dark matter density contrast is also recovered in the required amount. It should be noticed that the results (about growing density perturbations) are independent of the parameters of potential (\ref{cosh}). Thus, the predicted angular and power spectrums in the linear regime are those already shown in\cite{luis}. These spectrums are subject only to the imprint of the scalar potential (\ref{sinh}).

We also find an important difference with respect to the standard dark matter model. Analyzing the fluctuation equations, we clarified the meaning of a Jeans length for this model: it is related to the mass of scalar particle and to the time when scalar oscillations start. This Jeans lenght provokes the suppression in the power spectrum for small scales, that could explain the smooth core density of galaxies and the dearth of dwarf galaxies. Up to this point, the model has only one free parameter, $\lambda$. However, if we suppose that the scale of suppression is $k=4.5 h \, {\rm Mpc}$, then $\lambda \approx 20.28$, and then all parameters are completely fixed. From this, we found that $V_0 \simeq (36.5\, {\rm eV} )^4$ and the mass of the ultra-light scalar particle is $ m_\Phi \simeq 1.1\times 10^{-23}\, {\rm eV}$. The quantity $(\kappa _{o}V_0)^{-1/2}$ or the scalar mass $m_\Phi$ could play an important role in galaxies, appearing maybe in the observed constant core density of dark halos (see \cite{luis2,vlad}). This last fact could be a signature of the parameters of the dark matter potential (\ref{cosh}) as it has been shown that an exponential potential appears when analyzing scalar dark matter at galactic level\cite{siddh}. Further investigation will be published elsewhere.

A question could arise here: How good is eq. (\ref{ratio})?. Why does $\Omega_{0\gamma}$ also appear?. A possible answer could be that, in fact, potential (\ref{cosh}) can be written

\begin{equation}
V_0 \left[ \cosh{\left( \lambda \, \sqrt{\kappa_0} \Phi \right)} - 1 \right] = 2 V_0 \left[ \sinh{\ \left( \frac{\lambda}{2} \, \sqrt{\kappa_0} \Phi \right)} \right]^2.
\end{equation}

\noindent Then, potential (\ref{cosh}) is another sinh-like potential. Observe the similarity between eq. (\ref{ratio}) and the last of eqs. (\ref{tracpsi}), being the last one a generic solution of sinh-like potentials\cite{chimen}: They both involved the previous dominant component. And it is this singular feature that make us think about a cosmic coincidence of matter. Therefore, it would be clear that after a radiation-dominated era, a matter-dominated era must appear. Moreover, after a matter-dominated era, a dark energy-dominated era must appear, too. 

Summarizing, a model for the Universe where the dark matter and energy are of scalar nature can be realistic and could explain most of the observed structures and features of the Universe.

\acknowledgements{We would like to thank Vladimir Avila-Reese, F. Siddhartha Guzm\'an and Dar\'{\i}o N\'u\~nez for helpful discussions. This work was partly supported by CONACyT, M\'{e}xico 119259 (L.A.U.)}

\end{multicols}

\end{document}